\documentclass[twocolumn,aps,nofootinbib]{emulateapj}
\usepackage{amsfonts, amssymb}
\usepackage{graphicx, epsfig,bm}
\usepackage{color}
\def\stacksymbols #1#2#3#4{\def\theguybelow{#2}
        \def\verticalposition{\lower#3pt}
        \def\spacingwithinsymbol{\baselineskip0pt\lineskip#4pt}
        \mathrel{\mathpalette\intermediary#1}}
\def\intermediary #1#2{\verticalposition\vbox{\spacingwithinsymbol
        \everycr={}\tabskip0pt
        \halign{$\mathsurround0pt#1\hfil##\hfil$\crcr#2\crcr
                \theguybelow\crcr}}}

\shorttitle{Galaxy parameters estimation with Monte Carlo Markov Chains}
\shortauthors{Serra \it{et al.}}

\begin{document}

\title{CIGALEMC: Galaxy parameter estimation using a Markov Chain Monte Carlo 
approach with CIGALE}

\author{Paolo Serra\altaffilmark{1}, Alexandre Amblard\altaffilmark{1}, Pasquale Temi\altaffilmark{1}, Denis Burgarella\altaffilmark{2}, Elodie Giovannoli\altaffilmark{2},\\ Veronique Buat\altaffilmark{2}, 
Stefan Noll\altaffilmark{3}, Stephen Im\altaffilmark{1}}
\altaffiltext{1}{Astrophysics Branch, NASA/Ames Research Center, MS 245-6, Moffett Field, CA 94035.}
\altaffiltext{2}{Observatoire Astronomique de Marseille-Provence, 38 rue Frederic Joliot-Curie, 13388 Marseille Cedex 13, France.}
\altaffiltext{3}{Institut f\"ur Astro- und Teilchenphysik, Universit\"at Innsbruck, Technikerstr.25/8, 6020 Innsbruck, Austria }

\begin{abstract}
We introduce a fast Markov Chain Monte Carlo (MCMC) exploration of 
the astrophysical parameter space using a modified version of the 
publicly available code CIGALE (Code Investigating GALaxy emission). The 
original CIGALE builds a grid of theoretical Spectral Energy Distribution (SED) 
models and fits to photometric fluxes from Ultraviolet (UV) to 
Infrared (IR) to put contraints on parameters related to both 
formation and evolution of galaxies. Such a grid-based method can lead to 
a long and challenging parameter extraction since the computation time 
increases exponentially with the number of parameters considered and 
results can be dependent on the density of sampling points, which must be 
chosen in advance for each parameter. Markov Chain Monte Carlo methods, 
on the other hand, scale approximately linearly with the number of 
parameters, allowing a faster and more accurate 
exploration of the parameter space by using a smaller number of efficiently 
chosen samples. We test our MCMC version of the code CIGALE 
(called \texttt{CIGALEMC}) with simulated data. 
After checking the ability of the code to retrieve the input 
parameters used to build the mock sample, we fit theoretical SEDs to real 
data from the well known and studied SINGS sample. We discuss constraints 
on the parameters and show the advantages of our MCMC sampling method in terms of
accuracy of the results and optimization of CPU time.
\end{abstract}

\vskip.1in

\section{Introduction}
The spectral energy distribution (SED) of galaxies depends on many physical 
processes related to the emission from different 
stellar populations, absorption and re-emission from dust and gas and 
possible presence of Active Galactic Nuclei (AGN). 
Each process has been studied by many authors; libraries of stellar population 
models (Fioc \& Rocca-Volmerange (1997), Bruzual \&
Charlot (2003), Maraston (2005)), fitting curves for dust 
emission (Calzetti et al. (1994, 2000), Witt \& Gordon (2000)), 
studies of emission of dust grains (Chary \& Elbaz (2001), 
Dale \& Helou (2002), Lagache et al. (2003, 2004), and Siebenmorgen \&
Kr\"{u}gel (2007), Silva et al. (1998), Dopita et al. (2005), da Cunha et al. 
(2008)) are the basis of sophisticated fitting codes 
which derive physical parameters such as stellar mass, star formation rate, 
dust luminosity and so on.\\
Many parameters are usually necessary to describe these processes and 
model theoretical SEDs of galaxies. A grid of theoretical SED models 
is usually built and fitted to the 
data and statistical properties are derived for the parameters of interest.
A big drawback of any grid-based method is that, for any fitting process, 
the time to build models grows linearly with the number of models and 
then about exponentially with the number of parameters involved: 
such approaches are difficult to implement for complex models involving a 
sufficiently large number of parameters or when a fine 
sampling of the parameter space is necessary in order to retrieve 
statistically robust results. In the past few years, 
Markov Chains Monte Carlo (MCMC) techniques have started being widely 
used in physics. In cosmology, parameter estimation from cosmic microwave background data 
with MCMC methods has been introduced in Christensen et al. (2001) and 
has been implemented in the publicly available code 
\texttt{cosmomc} (Cosmological Monte Carlo, Lewis \& Bridle (2002))\footnote{\texttt{http://cosmologist.info/cosmomc/}};
in astrophysics, an MCMC approach to the stellar population syntesis modeling 
has been introduced in Conroy et al. (2009).\\
Here we use \texttt{cosmomc} as a 
generic sampler and we interface it to the publicly available code 
 CIGALE \footnote{\texttt{http://www.oamp.fr/cigale/}} (Code Investigation 
GALaxy Emission, Noll et al. (2009)) in order to allow a fast and accurate 
evaluation of the multidimensional parameter space probed by this code
\footnote{During the completion of this work we noticed that Acquaviva et al. (2011) have performed 
a similar work in the context of the code GALAXEV developed by 
Bruzual \& Charlot (2003).}. The main advantage of this method is that 
the computing time to fit the data scales
approximately linearly with the number of parameters involved, allowing 
the user to consider complex models with many parameters for only small 
additional computational time. MCMC techniques allow to probe also 
the shape of the probability distribution, giving far more information than just best 
fit and marginalized values for the parameters.\\
The paper is organized as follows; in the next section we briefly describe 
 CIGALE, introducing the main parameters used in the 
subsequent sections. We then explain the MCMC technique implemented 
in the modified version of CIGALE, 
which we call \texttt{CIGALEMC}. We test our code using a mock sample of 
$62$ galaxies already used in Giovannoli et at. (2011) and we apply it to a 
real galaxy sample with data from the Spitzer Infrared Nearby Galaxy 
Survey (SINGS, see Kennicutt et al. (2003)). We always consider a flat 
cosmological model with $\Omega_m=0.3$, $\Omega_{\Lambda}=0.7$ and $H_0=70\mathrm{km}\,\mathrm{s^{-1}}\mathrm{Mpc^{-1}}$. Finally we give our results 
and conclusions.

\section{The code CIGALE}
CIGALE calculates a grid of theoretical SEDs and fits to observational input data constitued by photometric filter fluxes ranging from 
UV to IR. For a detailed description of the code and its application 
to real data, we refer the interested reader to these papers 
(Burgarella et al. (2005), Noll et al. (2009), Giovannoli et al. (2011), Buat et al. (2011)).
In the following, we briefly summarize its main characteristics and 
the basic parameters used in the next sections. Our notation follows 
the one introduced in Giovannoli et al. (2011).

\subsection{Stellar populations and star formation rate} 
CIGALE combines both old and young stellar populations using 
single stellar populations of Maraston et al. (2005)
or PEGASE (Fioc \& Rocca-Volmerange (1997)). In this paper we will only 
use Maraston models; we assume star formation histories (SFH) with either 
exponentially decreasing star formation rate (SFR) in function of time t 
(``$\tau$ models''), as:  
  \begin{eqnarray}
\mathrm{SFR_{old}(t)}=\mathrm{SFR_{0,old}}\cdot\,e^{-\frac{(t-t_1)}{\tau_1}}\,\\
\nonumber
\mathrm{SFR_{young}(t)}=\mathrm{SFR_{0,young}}\cdot\,e^{-\frac{(t-t_2)}{\tau_2}}
\end{eqnarray}
or ``box models'' characterized by constant SFR over a limited period of 
time; in this case the instantaneous SFR at look-back time $\mathrm{t}'=0$ 
is given by the galaxy mass divided by the age t of the population, 
i.e. $\mathrm{M_{\mathrm{gal}}}/\mathrm{t}$.
Labels $1$ and $2$ refers to the old and young stellar populations 
while $\tau_{1,2}$ and $t_{1,2}$ (both is units of Gyr) are their e-folding 
time and age respectively. The two stellar populations are 
linked and weighted through their mass fraction; 
the parameter $f_{\mathrm{ySP}}$ represents the fraction of the 
young stellar mass over the total mass, so that the total instantaneous 
SFR (output parameter of CIGALE) is expressed as:
\begin{equation}
\mathrm{SFR=(1-f_{\mathrm{ySP}})SFR_{old}(t)+f_{ySP}SFR_{young}(t)}.
\end{equation}
\subsection{Absorption and emission by dust and gas}
In CIGALE, the absorption of star light by dust is described by a 
Calzetti attenuation curve (Calzetti et al. (1994, 2000)); possible modifications of the curve include both the addition of a UV bump at about $2175$\AA  (Fitzpatrick \& Massa (1990; 2007); Noll et al. (2009)) and the change of the slope through the multiplication by a power law $(\lambda/\lambda_V)^\delta$, where $\lambda_V=5500$\AA$\,$ is the reference wavelength for the $V$ filter; both the amplitude of the bump and the slope $\delta$ are free parameters of the code.
The attenuation correction is applied 
to both stellar populations individually using the visual attenuation parameter 
of the young stellar populations $(A_{\mathrm{ySP}}$, in units of magnitudes) and a reduction factor of the attenuation for the old model ($f_{V}$) as free parameters\footnote{The parameter $f_V $ has been labelled $f_{\mathrm{att}}$ in some previous papers.}. The code also calculates 3 additional parameters, derived from the final model SEDs; $A_{FUV}$ and $A_V$ are defined as the effective attenuation factors in magnitudes at $1500\pm 100$\AA$\,$  and $5500\pm100$\AA$\,$ while the age $\mathrm{t}_{\mathrm{D4000}}$ is derived from the D4000 break (see Balogh et al. (1999)) of the unreddened SED for a single stellar population\footnote{In its current version, CIGALE directly outputs the dust-free D4000 break, see Buat et al. (2011).}.\\
Dust emission in the IR is taken into account using 
64 templates of Dale \& Helou (2002). 
These models are parametrized by $\alpha$, 
the power law slope of the dust mass over heating intensity, 
defined as follows:
\begin{equation}
dM_d(U)=U^{-\alpha}dU,
\end{equation}
where $M_d(U)$ is the dust mass heated by a radiation field of intensity U.\\
Bolometric and dust luminosities ($\mathrm{L_{bol}}$ and $\mathrm{L_{dust}}$ respectively) are derived from all the basic parameters (see Noll et al. (2009)) and dust emission due to non-thermal sources such as AGN can also be added; 
the fraction $f_{\mathrm{AGN}}$ of dust luminosity $\mathrm{L_{dust}}$ (in $\mathrm{L_{\odot}}$) due to an AGN is estimated using AGN templates from Siebenmorgen et al. (2004).\\
The spectral line correction due to interstellar gas is performed as in 
Noll et al. (2009): for the optical band, empirical line templates are 
taken from the Kinney et al. (1996) starburst spectra while for the UV  
we use templates derived from SEDs presented in Noll et al. (2004). 
A correction for the redshift-dependent absorption of the intergalactic 
medium shortward of the Ly$\alpha$ line is also included using the algorithm of 
Meiksin (2006). 
\subsection{Comparison with data}
A grid of theoretical photometric fluxes is calculated at the redshift of the objects considered and a Bayesian analysis is performed through the calculation of the $\chi^2$ of each model:
\begin{equation}\label{eq_chi2}
\chi^2(\mathrm{M_\mathrm{gal}}) = \sum_{i=1}^{k} 
\frac{(M_\mathrm{gal} f_{\mathrm{mod},i} - f_{\mathrm{obs},i})^2}
{\sigma_{\mathrm{obs},i}^2}\equiv -2 \mathrm{ln}(\mathrm{L});
\end{equation}
here the galaxy mass $\mathrm{M}_\mathrm{gal}$ (in $M_\odot$) 
is treated as a free parameter, $f_{\mathrm{mod},i}$ and $f_{\mathrm{obs},i}$ are the theoretical and experimental 
fluxes respectively, the statistical photometry errors are considered in 
the term $\sigma_\mathrm{obs,i}$ and L is the normalized likelihood function.
\section{MCMC technique and \texttt{cosmomc}}
In Bayesian inference, the posterior probability of the parameters 
($\mathrm{\vec{\theta}}$) of a model in the light of the observed data 
($\mathrm{\vec{d}}$) is 
given by:
\begin{equation}
P(\vec{\theta}|\vec{d})=\frac{P(\vec{d}|\vec{\theta})P(\vec{\theta})}{P(\vec{d})};
\end{equation} 
here $\mathrm{P(\vec{d}|\vec{\theta})}\equiv\,\mathrm{L}(\vec{\theta})$ is the 
likelihood of the data given the model, $\mathrm{P(\vec{\theta})}$ is the 
prior on the parameters, which quantifies our {\it a priori} knowledge of the 
parameters and $\mathrm{P(\vec{d})}$ (called Evidence) is a normalization 
factor. In our case, $\mathrm{\vec{d}}$ represents the SED of each galaxy 
while $\mathrm{\vec{\theta}}$ represents the astrophysical parameters of CIGALE, 
as $\{\theta\}_i\equiv\,\{\tau_1,t_2,f_{ySP},...\}$. An MCMC sampler provides 
an efficient way 
to explore the posterior distribution and ensures that the number density 
of samples is asymptotically proportional 
to the probability density.
\begin{figure}[t!]
  \begin{center}
  \includegraphics[scale=0.6]{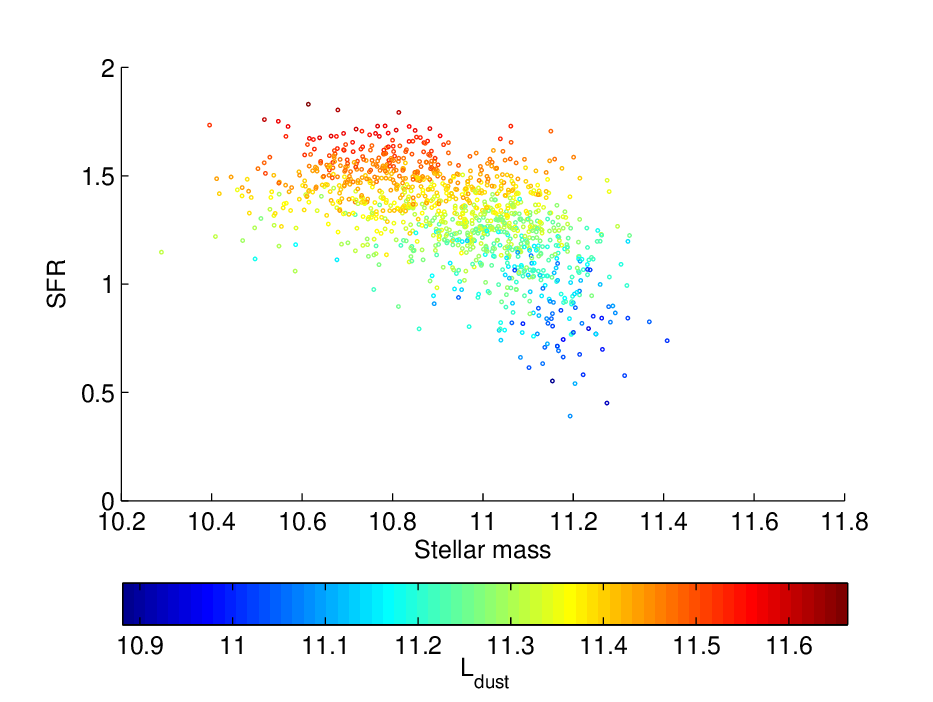}
    \caption{Samples from the posterior distribution for a test galaxy; the high density of points in the parameter space corresponds to large values of the posterior. Units are $\mathrm{M_{\odot}yr^{-1},\,M_{\odot},\,L_{\odot}}$ for $\mathrm{SFR,\,M_{star},\,L_{dust}}$ respectively.}
    \label{Figure1}
  \end{center}
\end{figure}
\subsection{Metropolis-Hastings algorithm}
The code \texttt{cosmomc} uses the Metropolis-Hastings 
algorithm to generate samples; each chain moves according 
to a transition probability $T(\vec{\theta}_i,\vec{\theta}_{i+1})$ 
which is determined so that the Markov Chain has a stationary asymptotic 
distribution equal to the posterior distribution $P(\vec{\theta})$ that we 
want to sample from. Given an arbitrary 
proposal density distribution ${q(\vec{\theta}_i,\vec{\theta}_{i+1})}$ to 
propose a new point $\vec{\theta}_{i+1}$ when the chain is at the 
point ${\vec{\theta}_{i}}$, the probability 
of transition is given by $\beta$:
\begin{equation}
\beta(\vec{\theta}_i,\vec{\theta}_{i+1})\,=\,\mathrm{min}\Big\{1,\frac{P(\vec{\theta}_{i+1})q(\vec{\theta}_{i+1},\vec{\theta}_i)}{P(\vec{\theta}_{i})q(\vec{\theta}_i,\vec{\theta}_{i+1})}\Big\}
\end{equation}
so that
\begin{equation}
T(\vec{\theta}_i,\vec{\theta}_{i+1})=\beta(\vec{\theta}_i,\vec{\theta}_{i+1})q(\vec{\theta}_i,\vec{\theta}_{i+1}).
\end{equation}
This ensures that the detailed balance holds:
\begin{equation}
P(\vec{\theta}_{i+1})T(\vec{\theta}_{i+1},\vec{\theta}_i)=P(\vec{\theta}_i)T(\vec{\theta}_i,\vec{\theta}_{i+1})
\end{equation}
and that the distribution converges to $\mathrm{P(\vec{\theta})}$. In practice, 
 a random number $x\in[0:1]$ is generated in the process of moving from $\vec{\theta}_i$ to $\vec{\theta}_{i+1}$ 
 so that the new point $\vec{\theta}_{i+1}$ is accepted if 
$\beta\geq\,x$. This ensures that each point of the chain depends only on 
its predecessor; in this sense the chain is a Monte-Carlo Markov process.

\subsection{Comparison with grid-based methods, {\it burn in} and 
convergence diagnostics}
As an illustration of the sampling mechanism, in Figure~1 we plot samples from the posterior distribution for a MCMC run with a test galaxy taken 
from a mock sample at redshift $z\sim0.7$ (see the following section for details); 
the number density of samples in the plane is proportional 
to the probability density of these two parameters. The dust luminosity $\mathrm{L_{\mathrm{dust}}}$ strongly depends on the SFR and the two parameters are degenerate, as shown by the colours in the figure. This 
plot clearly shows the efficiency of this MCMC method. In the grid-based 
approach, the parameter space is sampled in the same ``blind'' way for 
high and low values of the posterior: this can be an issue for both 
reliability of results and computation time, as also pointed out in Noll et al. (2009) and Acquaviva et al. 
(2011). In fact, local minima and degeneracies 
between parameters can be easily missed or undersampled without a good a 
priori knowledge of the parameter space; the oversampling of an 
ill-constrained parameter can also lead to a 
slight degradation of the estimates of well constrained parameters and  
many points can be generated in a region where the posterior is low, 
resulting in a waste of CPU time. This is not the case when MCMC chains are 
used because each chain ``learns'' where to move in the parameter 
space through the Metropolis-Hastings algorithm so that the density of 
samples is proportional to the posterior distribution. 
Degeneracies between parameters are more easily found, especially if 
many chains, starting from different 
regions in the parameter space, are used. In other words, 
the \texttt{CIGALEMC} user needs to specify the prior parameter space 
(number of parameters and their limits) but not the density of points for 
each parameter. In the following section we will provide a comparison 
of CPU time between the original CIGALE and \texttt{CIGALEMC} when 
evaluating physical properties of a mock sample. 
The code also calculates the covariance between various 
parameters so that an initial run can be made and the covariance matrix 
obtained can be used to improve the efficiency of sampling 
for subsequent runs.\\
Since each MCMC chain starts at a random position in the parameter space, 
it will take a little time before the chain equilibrates and starts 
sampling the posterior distribution. 
This period of initial convergence is called {\it burn in} period and 
the first {\it burn in} points of each chains will be discarded when doing any 
statistical analysis. In order to obtain uncorrelated samples of the 
posterior each chain is also ``thinned'' by using only occasional points of it; 
the {\it thinning} factor varies according to the number of parameters 
involved and it is tipically in the range 25-50. 
The code allows to choose the {\it burn in} fraction of the chain we want to 
discard and automatically thins out the chains.\\
In our analyses we won't use any initial covariance matrix for the parameters 
and, in order to be conservative, we compute statistical quantities using only 
the second half of each chain. In Figure~2 we plot the 
points of a MCMC chain for a galaxy sample in the plane ${\mathrm{L_{dust}}}$ vs 
$\mathrm{f}_{\mathrm{AGN}}$; the chain reaches the sensitive region 
of the parameter space after only a few ``burn in'' points characterized by
very low values of $\mathrm{L}_{\mathrm{dust}}$.   
Having a set of samples from the full posterior distribution, it is possible 
to calculate statistical quantities for the parameters of interest. 
Since the number density of sampling points is proportional to the 
posterior density, it is easy to calculate mean values and marginalized 
$1$-dimensional distribution for each parameter $\theta^i$ 
by simply counting the number N of samples within binned ranges of 
parameter values:
\begin{equation}
\langle\,\theta^i\rangle\sim\frac{1}{\mathrm{N}}\sum_{j=1}^N\theta^i_j, 
\end{equation}
while this is much more difficult in the context of the numerical grid 
integration because the calculation time grows exponentially with the 
number of dimensions.\\
In order to be sure that MCMC chains are efficiently sampling 
the posterior distribution (and then obtain robust statistics 
for each parameter) it is important to check their convergence. 
The code \texttt{cosmomc} provides 
two convergence criteria for runs with one single chain 
(Raftery \& Lewis,1992) and with multiple chains (Gelman \& Rubin, 1992). 
In the following analysis we will run multiple chains using the 
Gelman \& Rubin diagnostic which is 
characterized by the "variance of chain means"/"mean of chain 
variances" parameter R; $|\mathrm{R}-1| \leq 0.03$ is usually enough 
to reach convergence and stop the chains.\\ 
In this work we use \texttt{cosmomc} as a generic MCMC sampler and we link 
it to CIGALE in order to allow a faster exploration of the astrophysical 
parameter space. Our modified \texttt{CIGALEMC} code will be publicly available 
very soon\footnote{\texttt{http://www.oamp.fr/cigale/.}} and, 
since it is based on \texttt{cosmomc} for sampling options, convergence 
criteria and statistical quantities provided, we refer the reader to the  
website\footnote{\texttt{http://cosmologist.info/cosmomc/}} and to Lewis \& Bridle 
(2002) and references therein for a detailed explanation of the code and 
MCMC methods in general.

\begin{figure}[]
  \begin{center}
  \includegraphics[scale=0.8]{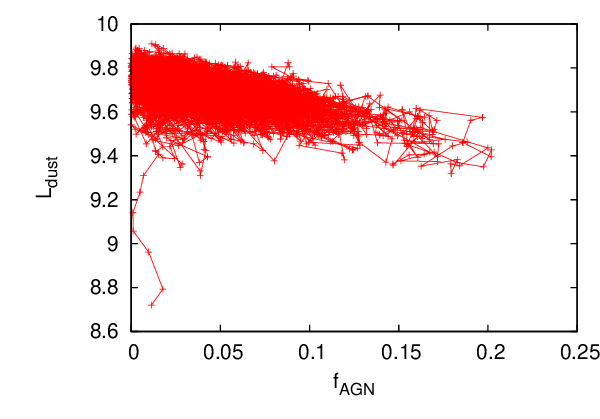}
    \caption{A Monte Carlo Markov chain in the 2-dimensional parameter space $\mathrm{L_{dust}}$ vs $f_{AGN}$. 
    The chain starts in a region where the likelihood is low (``burn in'' points with $\mathrm{L_{\mathrm{dust}}\sim8.8\,L_{\odot}}$) and quickly reaches the most sensitive region in the parameter space.}
    \label{Figure1}
  \end{center}
\end{figure}
\section{Analysis of a mock sample}
We test our \texttt{CIGALEMC} code with a mock sample already used in 
Giovannoli et al. (2011). We consider $62$ artificial galaxy SEDs 
corresponding to Luminous InfraRed Galaxies (LIRGs) at 
redshift $z\sim0.7$ and obtained by varying the input parameters of CIGALE in the following ranges:
 $[0.1\,\mathrm{Gyr}\le\tau_1\le10\,\mathrm{Gyr}, 0.025\,\mathrm{Gyr}\le t_2\le 0.7\,\mathrm{Gyr}, 0\le\,f_{ySP}\le\,1, 0.6\,\mathrm{mag}\le A_{ySP}\le 2.1\,\mathrm{mag},0\le f_{V}\le 1, 1.0\le\alpha\le 
2.5, 0\le f_{AGN}\le 0.3]$;
such a wide multidimensional parameter space allows to model very different 
spectral energy distributions of galaxies characterized by  
a wide range of possible star formation histories, 
absorption and emission by gas and dust and AGN contamination.
All galaxies are based on a Salpeter initial mass function; 
the age of the old stellar population is fixed at $7$ Gyr, we consider 
a constant star formation rate for the young population model ($\tau_2=20$ Gyr) 
and we do not add any modification to the original Calzetti attenuation 
curve (no UV bump and $\delta=0$).\\ 
Theoretical fluxes are calculated in the following $17$ bands from UV to IR:
$0.231\mu$m for GALEX, $[0.35-0.36-0.46-0.54-0.65-0.87-0.90-1.2-1.6-2.1]\mu$m 
(corresponding to MUSYC bands), $[3.6-4.5-5.8-8.0]\mu$m 
(corresponding to IRAC photometry) and $[24-70]\mu$m 
(corresponding to MIPS photometry). 
We add a gaussian distributed uncertainty $\mathrm{\sigma}$ to each 
theoretical flux; its value is $10\%$ of the corresponding flux, a reasonable choice for the measurements considered above.
For each galaxy we run 8 chains with initial positions randomly chosen  
in the parameter space determined by the following set of input parameters:
\begin{equation}
\{\mathrm{\tau_1,t_2,f_{ySP},A_{ySP},f_{V},\alpha,f_{AGN},M_{gal}}\}.
\end{equation} 
from which statistical quantities of interest can be calculated for the following set of derived parameters :
\begin{equation}
\{\mathrm{SFR,t_{D4000},L_{bol},L_{dust},A_{FUV},A_{V},M_{star}}\}.
\end{equation} 

We assume flat priors in the following parameters space: $[0.1\,\mathrm{Gyr}\le\tau_1\le10\,\mathrm{Gyr}, 0.025\,\mathrm{Gyr}\le t_2\le 2\,\mathrm{Gyr}, 0\le\,f_{ySP}\le1, 0\,\mathrm{mag}\le A_{ySP}\le 3\,\mathrm{mag},0\le f_{V}\le 1, 0.5\le\alpha\le 3, 0\le f_{AGN}\le 1,8\,\mathrm{M_{\odot}}<\mathrm{M_{gal}}<14\,\mathrm{M_{\odot}}]$; chains are stopped when the Gelman \& Rubin R-1 
parameter is  $|\mathrm{R}-1|\sim 0.03$.\\
First of all, we want to check that our results are 
statistically in agreement with the input values for the mock sample. 
As a tool derived from \texttt{cosmomc}, \texttt{CIGALEMC} allows to calculate and plot the mean likelihood and marginalized distribution for each parameter.
\begin{figure*}
  \begin{center}
    \begin{tabular}{ccc}
      \resizebox{60mm}{!}{\includegraphics{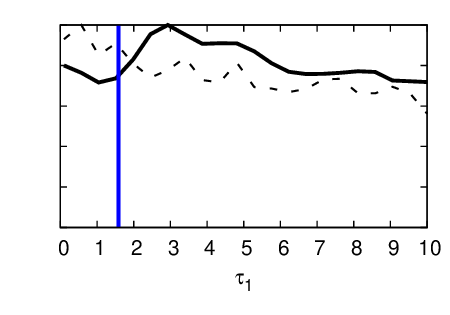}} &
      \resizebox{60mm}{!}{\includegraphics{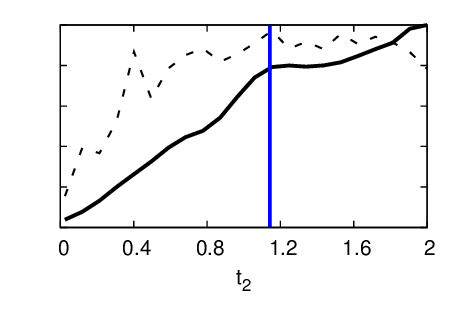}} &
      \resizebox{60mm}{!}{\includegraphics{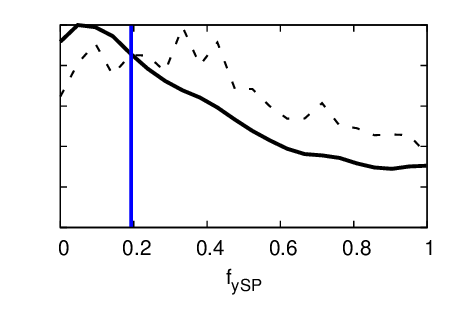}} \\
      \resizebox{60mm}{!}{\includegraphics{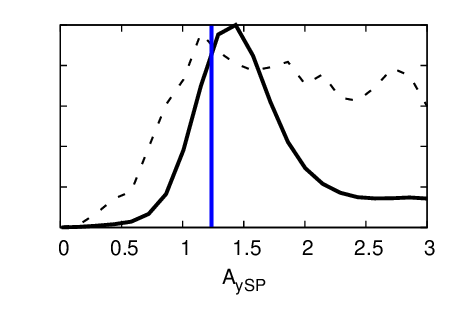}} & 
      \resizebox{60mm}{!}{\includegraphics{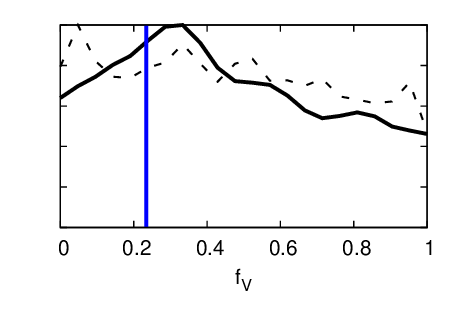}} &
      \resizebox{60mm}{!}{\includegraphics{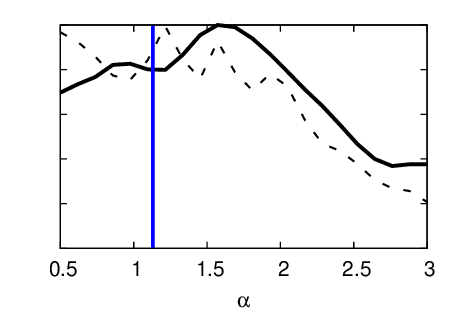}} \\
      \resizebox{60mm}{!}{\includegraphics{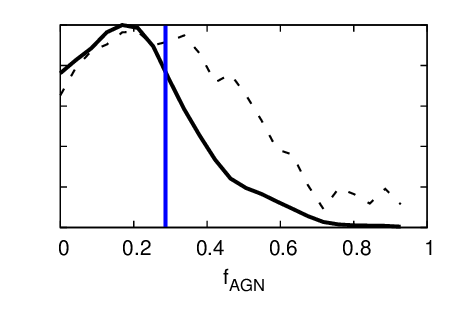}} &
      \resizebox{60mm}{!}{\includegraphics{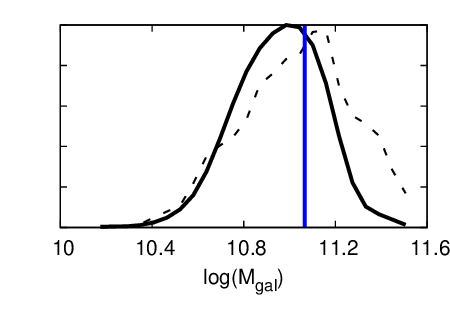}} &
      \resizebox{60mm}{!}{\includegraphics{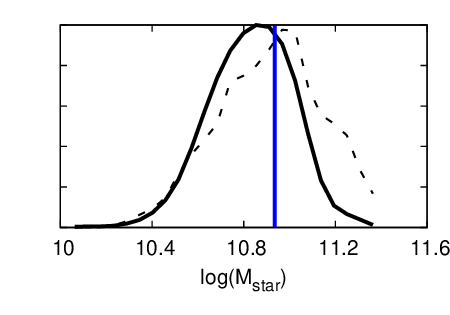}} \\
       \resizebox{60mm}{!}{\includegraphics{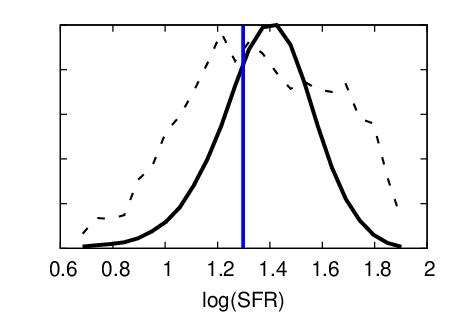}} &
        \resizebox{60mm}{!}{\includegraphics{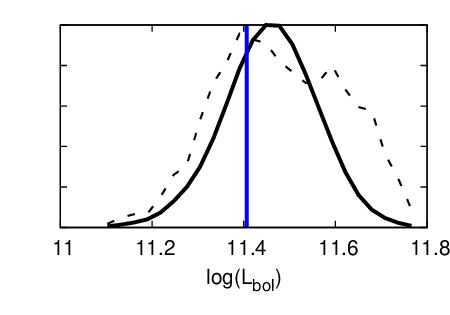}} &
        \resizebox{60mm}{!}{\includegraphics{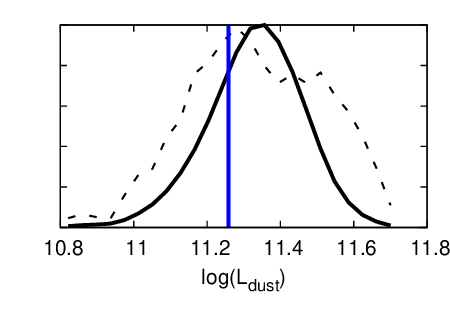}}\\
        \resizebox{60mm}{!}{\includegraphics{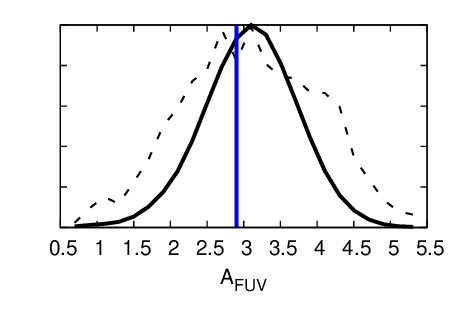}} &
        \resizebox{60mm}{!}{\includegraphics{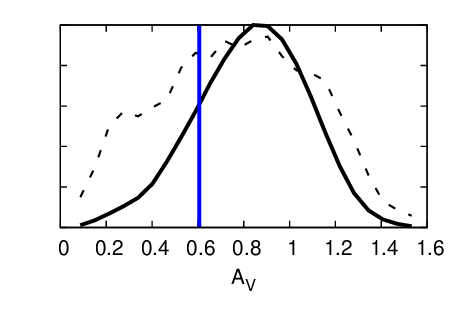}} &

    \end{tabular}
    \caption{Mean likelihoods, computed over the whole chains, for binned parameter values (dotted lines) and marginalized distributions (black lines) of some parameters of interest for an artificial galaxy of the mock sample considered; the blue vertical lines show the input parameter values used. Poorly constrained parameters, such as $\tau_1$, $t_2$, $f_{ySP}$, $f_{V}$ and $\alpha$ are clearly visible. Luminosities and masses are in units of solar luminosities and solar masses respectively.}
    \label{plotmock}
  \end{center}
\end{figure*}

The marginalized distribution in a given direction of the parameter space 
$\vec{d}={\bf h}(\vec{\theta})$ (where ${\bf h}(\vec{\theta})$ is the 
projector operator in one of the parameters considered, 
as ${\bf h}(\vec{\theta})=\theta_i$) is proportional to the 
number of samples at $\vec{d}$ and it can be 
expressed as:
\begin{equation}
\label{marge}
P(\vec{v}) = M(P,\vec{v}) \equiv \int d\vec{\theta} \, P(\vec{\theta}) \delta( {\bf h}(\vec{\theta}) - \vec{v}),
\end{equation}  
where $P(\vec{\theta})$ is the posterior distribution. Assuming flat priors on 
$\vec{\theta}$, the mean likelihood of samples with 
$\vec{d}={\bf h}(\vec{\theta})$ can be expressed as: 
\begin{equation}
\label{mean}
\langle\,P(\vec{v}:{\bf h}(\vec{\theta})=\vec{v})\rangle \equiv \frac{\int d\vec{\theta}\, P(\vec{\theta})^2 \delta({\bf h}(\vec{\theta}) - \vec{v})}{\int d\vec{\theta} \,P(\vec{\theta}) \delta({\bf h}(\vec{\theta}) - \vec{v})} = \frac{M(P^2,\vec{v})}{M(P,\vec{v})}.
\end{equation}

\begin{figure*}
  \begin{center}
    \begin{tabular}{ccc}
      \resizebox{60mm}{!}{\includegraphics{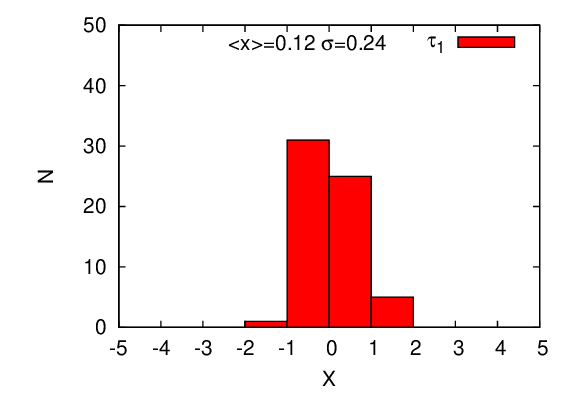}} &
      \resizebox{60mm}{!}{\includegraphics{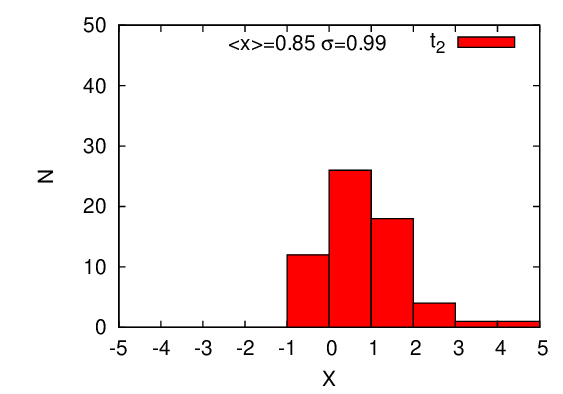}} &
      \resizebox{60mm}{!}{\includegraphics{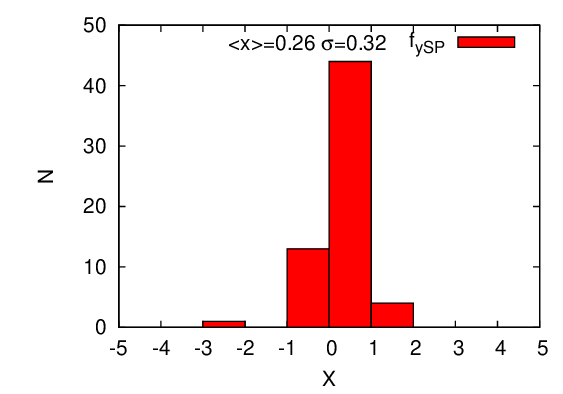}} \\
      \resizebox{60mm}{!}{\includegraphics{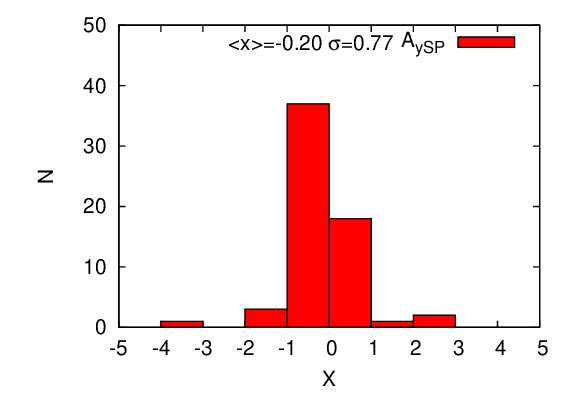}} & 
      \resizebox{60mm}{!}{\includegraphics{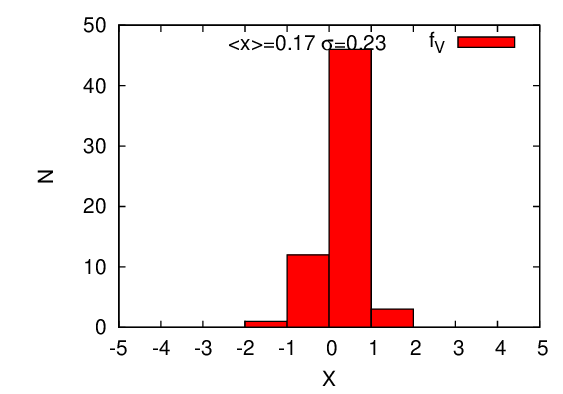}} &
      \resizebox{60mm}{!}{\includegraphics{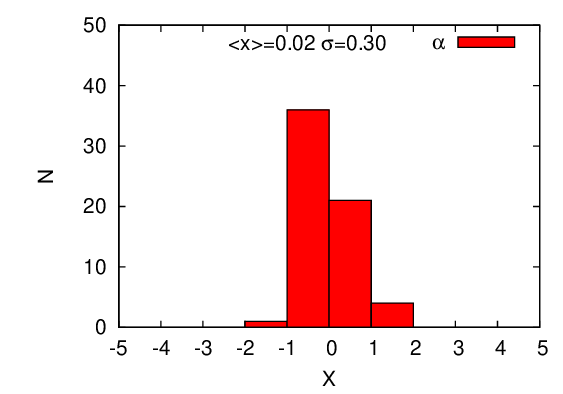}} \\
      \resizebox{60mm}{!}{\includegraphics{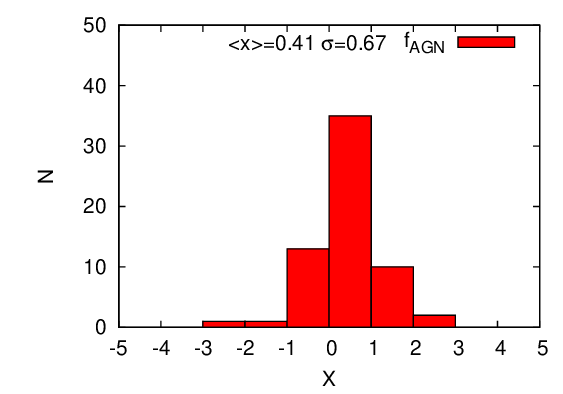}} &
      \resizebox{60mm}{!}{\includegraphics{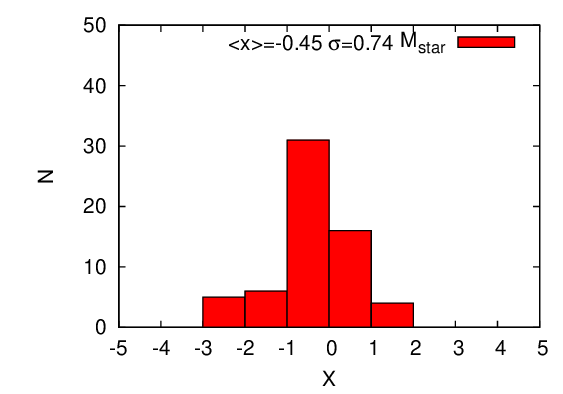}} &
      \resizebox{60mm}{!}{\includegraphics{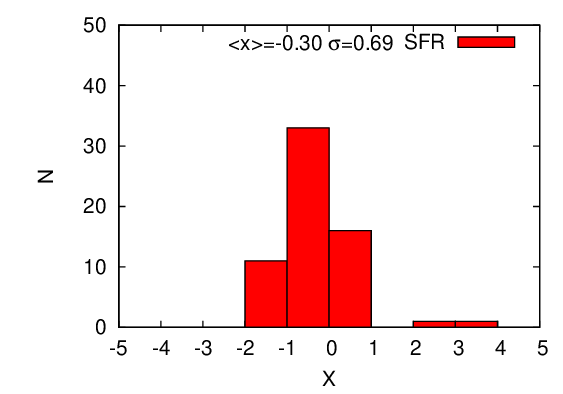}} \\
       \resizebox{60mm}{!}{\includegraphics{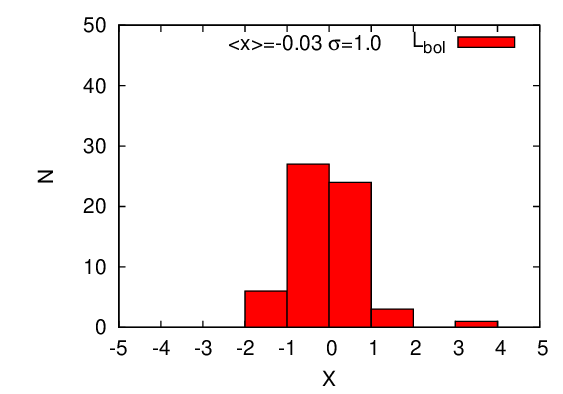}} &
        \resizebox{60mm}{!}{\includegraphics{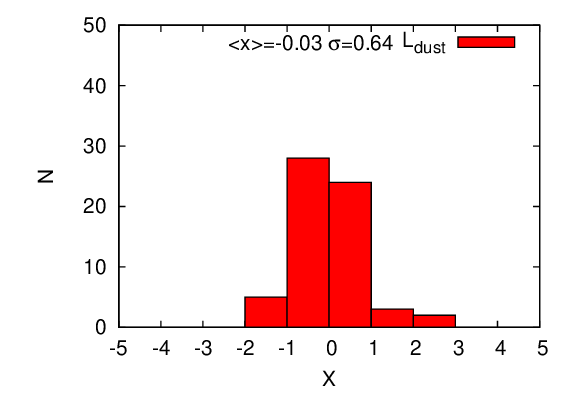}} &
        \resizebox{60mm}{!}{\includegraphics{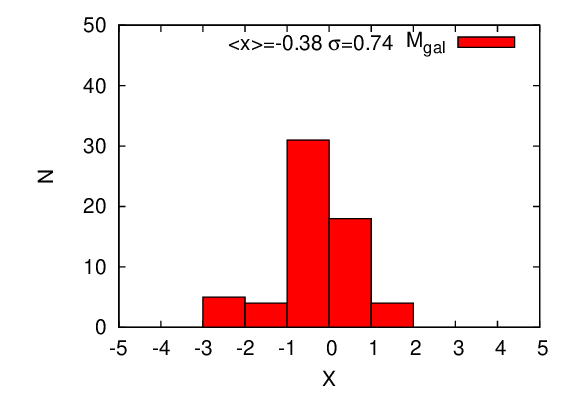}}\\
    \end{tabular}
    \caption{Distribution of the variable X for some parameters considered; X values are compatible with 0 at $68\%$ c.l..}
    \label{plotmock}
  \end{center}
\end{figure*}
If $P(\vec{\theta})$ is a multivariate Gaussian distribution it is possible 
to demonstrate that both mean likelihood and marginalized distribution are 
Gaussian and proportional so that they look the same: 
differences in these distributions will be then a signal of 
non-Gaussianity which can be either intrinsic or due to parameters not very 
well constrained.
In Figure~3 we show, for an artificial galaxy of the mock sample considered, both marginalized 
distributions (black solid lines) and mean likelihood (dotted lines) 
for some parameters of interest: we can see that $\tau_1$, $t_2$, $f_{ySP}$
$f_{V}$ and $\alpha$ are not very well constrained by the code. Similar results have been found in the analyses by Noll et al. (2009), Giovannoli et al. (2011) and Buat et al. (2011). \\ 
In order to study the goodness of the fit in a quantitative way for the 
whole sample of galaxies we introduce the quantity:
\begin{equation}
\mathrm{X^{i}} \equiv\,\frac{1}{\mathrm{N}} \sum_{j}\frac{\mathrm{O^i_{j}-I^i_{j}}}{\sigma^i_{j}};
\end{equation}
here $i$ runs for all the parameters considered by the code (so that 
$\mathrm{\{X\}^i=\{\tau_1,t_2,f_{ySP},...\}}$), 
$j$ runs for the $N=62$ objects of the mock, $\mathrm{O^{i}}$ 
is the set of the best fit values obtained as output from 
\texttt{CIGALEMC}, $\mathrm{I^{i}}$ is the vector of input parameters and 
$\mathrm{\sigma^{i}}$ is the vector of 
the $68\%$ c.l. marginalized uncertainties. 
As we can see from Figure~4, all $X$ values are compatible with $X=0$, 
which means that the code is able to find the best fit values of the 
parameters with high confidence. 
However, some distributions are slightly 
skewed: this can be due to chains being stuck in local minima of 
the likelihood function 
so that, in some cases, the best fit found does not correspond to the 
input value. 
This situation is typical when parameters are not well constrained (as for $t_2$) or in the presence  
of strong degeneracies between parameters.
As an example, Figure~5
shows the $68\%$ and $95\%$ c.l. in the plane $f_{ySP}$ vs $t_2$ for one 
sample galaxy of the mock: as we can see, 
limits on these parameters are not very strong; 
we also notice a partial degeneracy for high values of $f_{ySP}$ which 
is due to the fact that these two parameters affect the galactic SFR 
in the same way since we consider constant SFR for the 
young model:
\begin{equation}
\mathrm{SFR}\sim \mathrm{f_{ySP}\frac{M_{gal}}{t_2}}.
\end{equation}

\begin{figure}[t!]
\begin{center}
  \includegraphics[scale=0.51]{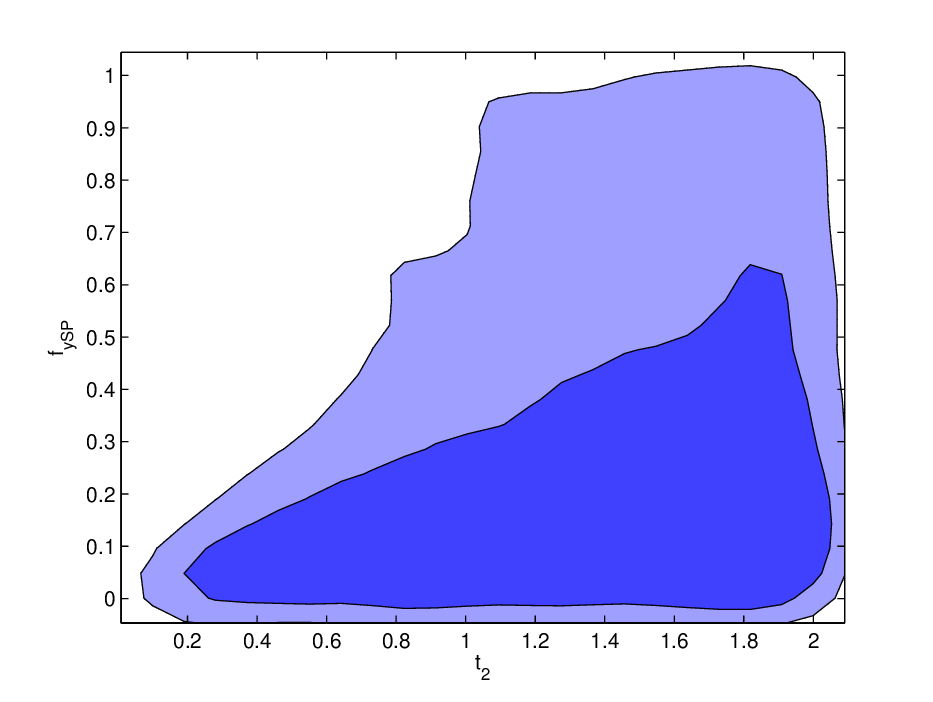} 
    \caption{Two-dimensional marginalized distribution showing the 
$68\%$ and $\%95$ c.l. 
contours for $f_{ySP}$ and $t_2$ for a galaxy sample;  
the parameter $t_2$ is unconstrained by the data and it is partially degenerate with $f_{ySP}$}
    \label{figure4}
\end{center}
\end{figure}

\begin{table}[h!t!]
\begin{center}
\begin{tabular}{|c|c|}
  \hline
 \textbf{Parameters}  & \textbf{mock sample} \\
  \hline
$\tau_{1}$ & 0.33 \\
log$_{10}~M_{\mathrm{star}}$ &   0.93    \\
log$_{10}~L_{\mathrm{bol}}$ & 0.92 \\
log$_{10}~L_{\mathrm{dust}}$   &   0.88   \\
log$_{10}~SFR$ & 0.81 \\
$t_2$ & 0.26 \\
 $f_{ySP}$ & 0.30  \\
 $A_{ySP}$ & 0.81 \\
 $f_{V}$ & 0.70   \\
  $\alpha$ & 0.65  \\
  $f_{AGN}$ & 0.71  \\
  \hline  
  \end{tabular}
  \caption{Estimation of the linear correlation coefficient of Pearson between the exact value and the value estimated by the code \texttt{CIGALEMC} for some parameters of the mock catalogue.}
  \label{table}
\end{center}
\end{table}
In order to check the reliability of our MCMC algorithm we tested the parameter uncertainty 
estimation. We created and fit the SEDs of $100$ artificial galaxies in 21 bands from IR to UV, built with the same input parameter set and with a $10\%$ scatter in flux, corresponding to the photometric error considered. We checked that in general we are able to find the true values within the region allowed at $68$ ($95$)$\%$ confidence $68$ ($95$)$\%$ of the time within the Poisson fluctuation error for different choices for both the input parameters and the prior ranges allowed in the fit.\\
Finally, we checked the consistency of our results by  
calculating the Pearson correlation coefficient $r$ between the 
input values of the parameters used to generate the mock and the best fit 
values found with \texttt{CIGALEMC}. The Pearson correlation 
coefficient quantifies the amount of correlation between two variables $X$ and $Y$ and it assumes values in the range $r\in[-1:1]$ (see Cohen (1988)). For samples ${X}_i$ and ${Y}_i$, it can be written as:
\begin{equation}
r=\frac{\sum_{i=1}^{N}(X_i-\bar{X})(Y_i-\bar{Y})}{(\sum_{i=1}^{N}(X_i-\bar{X})^2)^{\frac{1}{2}}(\sum_{i=1}^{N}(Y_i-\bar{Y})^2)^{\frac{1}{2}}};
\end{equation}
here $\bar{X}$ and $\bar{Y}$ denote the mean values of ${X}_i$ and ${Y}_i$.
The interpretation of the strength of the correlation can depend on the context; however, a widely used standard introduced by Cohen (1988) considers Pearson 
values $|r|>0.5$ as a signal of large correlation between variables, while $|r|<0.3$ denotes poor correlation.   
Again, as we can see from Table~1, some parameters ($\tau_1$, $t_2$, $f_{ySP}$) 
have small values of $r$; this is expected since these parameters are 
mostly unconstrained; however, even if the exact values of the input parameters are not found,  
they always fall inside the statistically significant region of the parameter space. We also explicitely checked that results do not significantly change when increasing the prior space for some parameters not very well constrained as $\mathrm{\tau_1}$, $\mathrm{t_2}$ and $\alpha$. Finally, we tested, for a galaxy of the mock, how results change when considering a gaussian prior, with different central values and width $\sigma=0.2$, on one of the most unconstrained parameters, $t_2$; we found that results on the 
other parameters are always statistically consistent respect to the choice of a flat prior on $t_2$.\\
In general, it is a good idea to choose the largest possible prior space for unconstrained parameters, 
in order to avoid possible biases due to the prior choice.\\
It is useful to compare the performance of CIGALE and \texttt{CIGALEMC} in terms of computing time, especially since computation can become prohibitive for any grid-based method if the number of parameters involved is sufficiently high. The CPU time required to obtain convergence of the chains for each galaxy 
mainly depends on both the quality of the data and the number of parameters 
considered. Running 8 chains in parallel (each one on a 2.40 GHz Intel Xeon E5530), for each galaxy of the mock we typically reach a good convergence with $\sim 20000$ points for each chain, which means a total of $\sim$ 160000 points. The grid built in Giovannoli et al. (2011) to analyze the same sample with the same number of free parameters and bounds contained $\sim3.5\cdot10^6$ points; this means a gain of 
order $\sim$20 in efficiency in the estimation of the parameters but a more dramatic efficiency can be easily reached when we need to use either more parameters or a fine sampling in a given direction of the parameter space or both. The average CPU time to reach a good convergence for a galaxy of this mock is about 35 seconds, which translates in $280$s of total CPU time.
\section{Analysis of real data: the SINGS sample}
We now use \texttt{CIGALEMC} to infer physical properties of the well known 
SINGS (Spitzer Infrared Nearby Galaxy Survey; Kennicutt et al. (2003)) 
sample. In order to make a comparison with results obtained using the
 grid-based CIGALE, we use the same 39 galaxies already considered in 
Noll et al. (2009) with the same spectral coverage: GALEX FUV ($\sim 1500$\AA) 
and NUV ($\sim 2300$\AA) filters (Gil de Paz et al. (2007)), 
2MASS data for J,H, K$_s$ (Jarrett et al. (2003)), IRAC and MIPS filters for 
$[3.6, 4.5, 5.8, 8.0, 24, 70, 160]\mu$m (Dale et al. (2005)), 
Dale et al. (2007, 2008) optical data for B, V, R and 
I bands corrected as in Mu\~noz-Mateos et al (2009) and fluxes 
from $u'$, $g'$, $r'$, $i'$ and $z'$ filters of SDSS 
(Stoughton et al. (2002)); Dale et al. (2007, 2008) optical data are only used for $14$ galaxies  
for which SDSS data are not available. The mean photometric relative uncertainties for the bands 
considered are shown in Table~2; in particular, very small uncertainties are associated with the 2MASS 
bands. We performed a preliminar run, realizing that the hard 
constraints coming from this filter set did not allow the code to properly fit for the other filters. 
Since these uncertainties do not take into account for possible calibration errors and 
since a systematic offset can also affect the CIGALE theoretical 
models, we decided to be conservative and, following Noll et al. (2009), we add a $5\%$ uncertainties 
in quadrature for each filter set.\\
 \begin{table}[h!t!]
\begin{center}
\begin{tabular}{|c|c|}
  \hline
 \textbf{Filters}  & \textbf{Rel. errors} \\
  \hline
GALEX FUV, NUV & 15$\%$ \\
Dale et al. B, V, R, I & 16$\%$ \\
SDSS uŒ , gŒ , rŒ , iŒ, zŒ &  $3\%$ \\
2MASS J, H, Ks & $1\%$ \\
IRAC 3.6, 4.5, 5.8, 8.0 $\mu\,m$ & $11\%$ \\
MIPS 24 $\mu$\,m & $5\%$ \\ 
MIPS 70 $\mu$\,& $7\%$ \\ 
MIPS 160 $\mu$\,m & $13\%$ \\
  \hline  
  \end{tabular}
  \caption{Mean photometric uncertainties for our SINGS sample.}
  \label{table}
\end{center}
\end{table}    
In our analysis we assume the following ange of variation for a set of 9 astrophysical parameters: 
$[0.1\,\mathrm{Gyr}\le\tau_1\le 10\,\mathrm{Gyr},0.025\,\mathrm{Gyr}\le t_2 \le 2\,\mathrm{Gyr}, 0 \le f_{ySP} \le 
1,-0.5 \le \delta \le 0.5, 0\,\mathrm{mag}\le A_{ySP}\le 5\,\mathrm{mag}, 0 \le f_{V} \le 1, 0.5\le\alpha\le 
3, 0\le f_{AGN}\le 1,8\mathrm{M_{\odot}},\mathrm{M_{gal}}<13\mathrm{M_{\odot}}]$. We keep fixed both the age of the old 
stellar population ($t_1=10$ Gyr) 
and the e-folding time for the young stellar population ($\tau_2=20$ Gyr): 
these parameters are not well constrained by the data so that fixing them 
does not alter the fit. Finally, We only consider models with  Salpeter initial 
mass function and solar metallicity; metallicity measurements are quite uncertain due to many limiting factors (Noll et al. (2009), Moustakas $\&$ Kennicutt (2006) and references therein); Noll et al. (2009) checked the influence of different assumptions for the metallicity, concluding that deviations in the properties of the galaxies are within the uncertainties, which tend to increase by $0$ to 20$\%$ when half or double of the solar metallicity are considered. The only exception is the absolute value of $t_{D4000}$, because of the well-known age-metallicity degeneracy (e.g., Kodama \& Arimoto, (1997)).\\ 
Our reference AGN model has $L=10^{12}L_{\odot}$, $R=125$pc and $A_V=32$ for the 
luminosity of the non-thermal source, the outer radius of a spherical dust 
cloud covering the AGN and the amount of attenuation in the optical 
caused by the cloud respectively.\\ 
Our findings can be summarized as follows:
\begin{itemize}
\setlength{\itemsep}{-0.8mm}
\item Very good constraints are derived for the AGN fraction of all sources ($f_{AGN}\le\,0.10$) except for NGC0584 and NGC1404 for which $\mathrm{f_{AGN}}\leq0.6$ at $68\%$ c.l.. We note that the flux at 160$\mu$m for NGC0584 can be contaminated by some foreground/background emission and the most recent Herschel data conclusively confirm that a background source contaminates both fluxes at 70 and 160$\mu$m for NGC1404 (Daniel Dale, private communication); the low quality of these data is responsible for the big uncertainties obtained for other parameters as $\mathrm{SFR\,L_{bol}\,L_{dust}\,A_{FUV}}$; in Figure~6 we plot 
the 2-dimensional marginalized distribution for SFR vs $f_{AGN}$ for NGC1404: the double-peaked likelihood function is most probably an artifact due to the low quality of data for this source. In general, we note that that NGC0584 and NGC1404 are elliptical galaxies which tend to have very weak but warm dust emission; it is not surprising that they show an apparently high AGN fraction. We decided not to use these sources in the rest of our analysis.
\item We are not able to put strong constraints on ``phenomenological'' 
parameters as $\tau_1$, $t_2$ and $\delta$; limits on these parameters 
depend on the assumed prior range. 
The poor determination of $\delta$ is essentially due to the 
low number of data in UV. In general, both $\delta$ and the possible UV bump 
allowed by the code are very difficult to constrain with few only broad band 
data (Buat et al. (2011) and Buat et al., in preparation). The fraction of the young stellar mass over the total mass is well constrained, with values $f_{ySP}<\,0.26$ for all the galaxies, except for NGC1705, NGC2798 and NGC4631, for which $f_{ySP}$ is unconstrained. The mean value for the parameter $\alpha$ of the Dale $\&$ Helou (2002) templates is $\alpha=2.44$ (in agreement with Noll et al. (2009); we note that, while low values of $\alpha$ are well constrained, no constraints are found on high values, with $\alpha=3$ compatible at 68$\%$ c.l. for all galaxies expect NGC2798, for which $\alpha<1.77$ at 68$\%$ c.l.; weak constraints on high values for $\alpha$ are due to the degeneracy of the dust emission models for wavelengths shortwards of the emission peak, so that, for $\alpha>\,2.5$, the flux ratio $\mathrm{f_{\nu}(60)}\mathrm{\mu}m/\mathrm{f_{\nu}(100)}\mathrm{\mu}m$ is almost constant (see Noll et al. (2009).
\item Good constraints can be derived for the mass dependent parameters 
$\mathrm{M_{star}}$, $\mathrm{SFR}$, $\mathrm{L_{bol}}$ and $\mathrm{L_{dust}}$ 
as shown in Table~3. The Pearson correlation coefficient with results from Noll et al. (2009) is $r=0.99,\,0.98,\,0.94,\,1.0,\,1.0,\,0.94$ for $\mathrm{M_{star},\,SFR,\,tD4000,\,L_{bol},\,L_{dust},\,A_{FUV}}$ respectively. In Figure 8 we show a quantitative comparison with the analysis performed by Noll et al. (2009) with the original code CIGALE by plotting the ratio, $\mathrm{Q}\equiv(\mathrm{\texttt{CIGALEMC}-\mathrm{CIGALE}})/\sigma_{\texttt{CIGALEMC}}$, where \texttt{CIGALEMC} and CIGALE refer to the mean value of the parameters quoted in Table~3 for this work and Noll et al. (2009) respectively. Possible systematic differences between the results of both methods can be studied by considering mean, standard deviation and skewness for the parameters of interest. As we can see from Table~4, no significant difference between results in this paper and in Noll et al. (2009) is found for $\mathrm{SFR}$, $\mathrm{log(t_{D4000})}$, $\mathrm{log(\mathrm{L}_{\mathrm{bol}})}$, and $\mathrm{A_{\mathrm{FUV}}}$, with each value compatible with 0 at less or about $68\%$ c.l.; however, there is some skewness associated to $\mathrm{SFR}$, $\mathrm{t_{D4000}}$ and $L_{bol}$. In case of $\mathrm{M_{star}}$ and $\mathrm{L_{dust}}$, the mean values are lower from 0 at about 95$\%$ c.l., indicating the existence of some offset between results in this paper and in Noll et al. (2009). A further comparison of the uncertainties between our results and Noll et al. (2009) shows that we are able to put stronger constraints on the galaxy SFR, while our constraints on $\mathrm{M_{star},\,L_{bol},\,L_{dust},\,A_{FUV}}$ are weaker. Differences between our results and results in Noll et al. (2009) can be due to the different choice of the input parameter space in terms of both number and type of parameters used in the analysis; in particular, we consider a wider range of variation for $\mathrm{\alpha}$, $\mathrm{\delta}$, $\mathrm{A_{ySP}}$, $\mathrm{t_2}$ and $\mathrm{f_{AGN}}$, while we decide to fix the parameter $\mathrm{\tau_2}$ (see Table 3 of Noll et al. (2009) for their choice of the parameter space); however, it is reassuring to see how different theoretical assumptions lead to compatible results.
\item A frequency dependent $\chi^2$ analysis shows which bands mostly contribute to the total $\chi^2$ for each galaxy. 
We introduce the averaged $\mathrm{\chi^2}$:
\begin{equation}
\mathrm{\langle\chi^2(\lambda^j)\rangle=\frac{1}{N_j}\sum_{i=1}^{N_j}\Big(\frac{Th_i^j-d_i^j}{\sigma^j_i}\Big)^2};
\end{equation}      
here $\mathrm{Th^j_i}$ and $\mathrm{d^j_i}$ are respectively the theoretical best fit SEDs and the data points for the $i$-th galaxies at the $j$-th 
frequency and the sum runs over our sample of galaxies. From Figure~7 we see that the code is able to find a good agreement with the data for all the bands considered. The frequency dependent $\chi^2$ is particularly low for bands with 
large relative errors as UV GALEX bands and optical B, V, R, I bands. The large uncertainty found for the MIPS $70\mu$m filter is mainly due to the galaxies NGC1715 and NGC5866; the code is not able to find a good fit for the $70\mu$m filter and the reduced chi-squared is high ($\chi^2/d.o.f=7.8$ for both galaxies, see Table 3). A previous analysis by Cannon et al. (2006), based on Spitzer observations, found similar results for NGC1705, showing in particular that the models of Li $\&$ Draine (2001, 2002) give a better fit to IR data than the Dale \& Helou (2002) models used in this paper (see Figure~3 of Cannon et al. (2006)). Interestingly enough, NGC1715 and  NGC5866 are, respectively, the only irregular and S0 galaxies in our sample.\\ 
In general, we find a small trend of worse fitting for galaxies with small SFR values. Galaxies with the lowest values of the $\chi^2$ do not show peculiar properties, with total stellar masses of about $10^{11}\mathrm{M_{\odot}}$ and SFRs between 0.1 and 10$ \mathrm{M_{\odot}}/yr$, typical of nearby spiral galaxies; an exception is the dwarf galaxy NGC4625 (reduced $\chi^2/d.o.f=1.2$), characterized by smaller values for both the stellar mass and luminosity. Three other dwarf galaxies (NGC1705, NGC2976 and NGC5474) are clearly identified in our sample by looking at their smaller values for both the luminosity and the stellar masses respect to the rest of the sample.
Finally, in order to check for a possible correlation between the filter set used and the goodness of the fit, we also calculated the mean $\bar{\chi^2}$ for the subset of galaxies for which  SDSS filters are available respect to the subset of galaxies with optical Dale et al. filters; we found no correlation, with $\bar{\chi}\sim 2.5$ in both cases.
              
\end{itemize}

\begin{figure}[t!]
\begin{center}
  \includegraphics[scale=0.51]{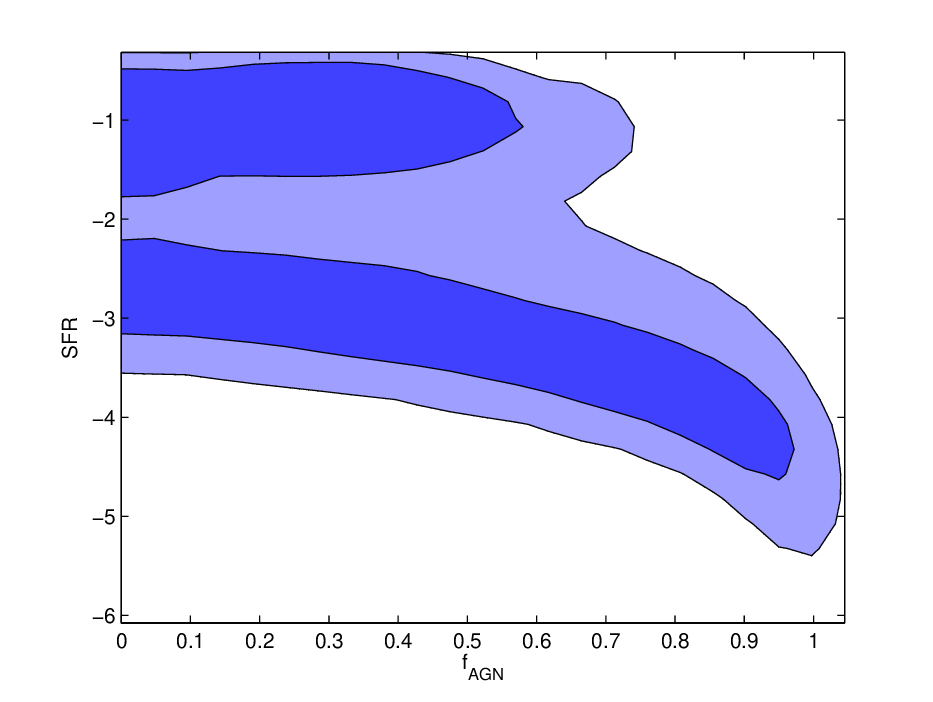} 
    \caption{Two-dimensional marginalized distribution showing the $68\%$ and $\%95$ c.l. contours for $\mathrm{log(SFR)}$ vs $f_{AGN}$ for NGC1404; 
the double peaked likelihood is clearly visible and $f_{AGN}$ is mostly unconstrained.}
    \label{figure5}
\end{center}
\end{figure}

\begin{table*}[tl!]
\caption{Mean values and 68$\%$ c.l. marginalized results for some parameters related to the SINGS sample considered.}
\label{tab_SINGS}
\centering
\begin{tabular}{l l  r@{\ $\pm$\ }l r@{\ $\pm$\ }l r@{\ $\pm$\ }l
r@{\ $\pm$\ }l r@{\ $\pm$\ }l r@{\ $\pm$\ }l  lr}
\hline\hline
\noalign{\smallskip}
ID & Type & 
\multicolumn{2}{c}{$\log M_\mathrm{star}$} & 
\multicolumn{2}{c}{$\log \mathrm{SFR}$} &
\multicolumn{2}{c}{$\log t_\mathrm{D4000}$} &  
\multicolumn{2}{c}{$\log L_\mathrm{bol}$} &  
\multicolumn{2}{c}{$\log L_\mathrm{dust}$} &
\multicolumn{2}{c}{$A_\mathrm{FUV}$} & 
\multicolumn{2}{c}{$\chi^2$/d.o.f.} \\ 
    \multicolumn{2}{c}{} & 
\multicolumn{2}{c}{[M$_{\odot}$]} & 
\multicolumn{2}{c}{[M$_{\odot}$/yr]} &
\multicolumn{2}{c}{[Gyr]} &
\multicolumn{2}{c}{[L$_{\odot}$]} & 
\multicolumn{2}{c}{[L$_{\odot}$]} & 
\multicolumn{2}{c}{[mag]} &
\multicolumn{2}{c}{} \\
\noalign{\smallskip}
\hline
\noalign{\smallskip}
NGC\,0024& SAc  & 9.52 & 0.10 & -0.78 & 0.11 & -0.14 & 0.18 & 9.46 & 0.05 & 8.62 & 0.11 & 0.45 & 0.16 & 2.5 &\\
NGC\,0584&  E4 & 11.41 & 0.04 & -1.3 & 0.25 & 0.99 & 0.02 & 11.09 & 0.18 & 10.22 & 1.14 & 4.47 & 4.41 & 2.7 &\\
NGC\,0925&  SABd & 9.93 & 0.17 & 0.18 & 0.12 & -0.43 & 0.10 & 10.24 & 0.07 & 9.64 & 0.09 & 0.6 & 0.21 & 2.6 &\\
NGC\,1097& SBb  & 11.19 & 0.11 & 0.9 & 0.11 & -0.15 & 0.17 & 11.15 & 0.04 & 10.77 & 0.08 & 1.85 & 0.51 &0.5 & \\
NGC\,1291& SBa  & 11.14 & 0.05 & -0.72 & 0.23 & 0.86 & 0.08 & 10.68 & 0.04 & 9.24 & 0.16 & 0.98 & 0.66 &1.9 & \\
NGC\,1316& SAB0 & 12.01 & 0.05 & -0.06 & 0.28 & 0.91 & 0.07 & 11.54 & 0.04 & 10.07 & 0.12 & 1.49 & 1.07 &4.5& \\
NGC\,1404& E1  & 11.52 & 0.04 & -1.0 & 0.20 & 0.98 & 0.02 & 11.25 & 0.21 & 10.76 & 0.58 & 5.58 & 5.07 & 1.4 &\\
NGC\,1512&  SBab & 10.34 & 0.09 & -0.28 & 0.14 & 0.17 & 0.32 & 10.13 & 0.04 & 9.39 & 0.09 & 0.86 & 0.30 & 1.8 &\\
NGC\,1566& SABbc & 10.88 & 0.11 & 0.9 & 0.10 & -0.34 & 0.08 & 11.02 & 0.05 & 10.58 & 0.08 & 1.13 & 0.34 & 1.6 &\\
NGC\,1705& Am  & 8.20 & 0.20 & -1.15 & 0.15 & -0.62 & 0.25 & 8.81 & 0.10 & 7.67 & 0.12 & 0.11 & 0.05 & 7.8 &\\
NGC\,2798& SBa & 10.03 & 0.18 & 0.61 & 0.07 & -0.54 & 0.06 & 10.6 & 0.05 & 10.47 & 0.06 & 4.48 & 0.64 & 2.0 &\\
NGC\,2841&  SAb & 10.92 & 0.06 & -0.3 & 0.17 & 0.62 & 0.10 & 10.54 & 0.03 & 9.57 & 0.11 & 1.04 & 0.46 & 1.3 &\\
NGC\,2976& SAc & 9.33 & 0.09 & -0.8 & 0.08 & -0.28 & 0.06 & 9.37 & 0.04 & 8.87 & 0.09 & 1.11 & 0.28 & 3.4 &\\
NGC\,3031& SAab  & 10.96 & 0.06 & -0.16 & 0.14 & 0.56 & 0.10 & 10.59 & 0.03 & 9.51 & 0.11 & 0.63 & 0.27 & 1.8 &\\
NGC\,3184& SABcd  & 10.14 & 0.08 & 0.11 & 0.07 & -0.33 & 0.07 & 10.24 & 0.03 & 9.68 & 0.09 & 0.78 & 0.23 & 4.4 &\\
NGC\,3190& SAap  & 10.87 & 0.04 & -0.65 & 0.35 & 0.75 & 0.15 & 10.44 & 0.03 & 9.62 & 0.10 & 3.16 & 1.23 & 1.4 &\\
NGC\,3198& SBc   & 10.01 & 0.08 & -0.02 & 0.07 & -0.33 & 0.07 & 10.12 & 0.04 & 9.55 & 0.09 & 0.73 & 0.21 & 2.2 &\\
NGC\,3351& SBb & 10.58 & 0.08 & -0.02 & 0.12 & 0.14 & 0.23 & 10.38 & 0.04 & 9.82 & 0.09 & 1.36 & 0.41 & 1.1 &\\
NGC\,3521& SABbc & 10.95 & 0.08 & 0.41 & 0.14 & 0.06 & 0.23 & 10.76 & 0.04 & 10.28 & 0.10 & 2.06 & 0.51 & 0.9 &\\
NGC\,3621& SAd & 10.04 & 0.12 & 0.14 & 0.10 & -0.38 & 0.08 & 10.24 & 0.05 & 9.82 & 0.09 & 1.15 & 0.35 & 1.4 &\\
NGC\,3627& SABb & 10.80 & 0.10 & 0.55 & 0.11 & -0.21 & 0.08 & 10.77 & 0.04 & 10.38 & 0.09 & 2.0 & 0.40 & 1.7 &\\
NGC\,4536& SABbc & 10.89 & 0.12 & 1.0 & 0.10 & -0.39 & 0.08 & 11.11 & 0.04 & 10.79 & 0.08 & 1.65 & 0.40 & 1.2 &\\
NGC\,4559& SABcd  & 10.11 & 0.16 & 0.43 & 0.08 & -0.47 & 0.06 & 10.46 & 0.04 & 9.84 & 0.09 & 0.52 & 0.15 & 3.3 &\\
NGC\,4569& SABab & 11.38 & 0.05 & 0.33 & 0.17 & 0.53 & 0.12 & 11.03 & 0.03 & 10.26 & 0.10 & 1.66 & 0.56 & 4.0 &\\
NGC\,4579& SABb & 11.42 & 0.06 & 0.18 & 0.21 & 0.63 & 0.11 & 11.03 & 0.03 & 10.13 & 0.11 & 1.5 & 0.60 & 1.3 &\\
NGC\,4594& SAa & 11.73 & 0.05 & -0.44 & 0.27 & 0.94 & 0.06 & 11.26 & 0.04 & 9.7 & 0.12 & 1.29 & 0.97 &  1.4 & \\
NGC\,4625& SABmp  & 9.18 & 0.10 & -0.77 & 0.08 & -0.37 & 0.08 & 9.33 & 0.04 & 8.79 & 0.10 & 0.75 & 0.22 & 1.2 &\\
NGC\,4631& SBd  & 10.08 & 0.17 & 0.76 & 0.07 & -0.59 & 0.06 & 10.73 & 0.04 & 10.45 & 0.08 & 1.39 & 0.36 & 2.9 &\\
NGC\,4725& SABab & 11.34 & 0.07 & 0.31 & 0.15 & 0.51 & 0.11 & 11.0 & 0.03 & 10.01 & 0.12 & 0.7 & 0.29 & 2.3 &\\
NGC\,4736& SAab  & 10.75 & 0.07 & 0.08 & 0.12 & 0.21 & 0.30 & 10.51 & 0.03 & 9.84 & 0.10 & 1.2 & 0.36 & 1.7 & \\
NGC\,4826 & SAab& 10.77 & 0.05 & -0.44 & 0.21 & 0.62 & 0.12 & 10.38 & 0.03 & 9.54 & 0.11 & 1.78 & 0.68 & 2.6 &\\
NGC\,5033& SAc & 10.68 & 0.10 & 0.43 & 0.11 & -0.20 & 0.09 & 10.65 & 0.04 & 10.25 & 0.09 & 1.7 & 0.42 & 1.2 &\\
NGC\,5055& SAbc & 10.88 & 0.08 & 0.45 & 0.11 & -0.06 & 0.14 & 10.75 & 0.04 & 10.27 & 0.09 & 1.7 & 0.41 & 2.2  &\\
NGC\,5194& SABbc & 10.74 & 0.11 & 0.85 & 0.09 & -0.39 & 0.08 & 10.94 & 0.04 & 10.55 & 0.09 & 1.3 & 0.33 & 0.8 & \\
NGC\,5195& SB0p & 10.76 & 0.04 & -0.42 & 0.27 & 0.61 & 0.15 & 10.38 & 0.04 & 9.61 & 0.15 & 2.55 & 0.92 & 5.6 &\\ 
NGC\,5474& SAcd  & 9.22 & 0.14 & -0.53 & 0.10 & -0.43 & 0.07 & 9.52 & 0.04 & 8.52 & 0.12 & 0.19 & 0.07 & 5.6 &\\
NGC\,5713& SABbcp & 10.54 & 0.19 & 0.83 & 0.11 & -0.44 & 0.09 & 10.89 & 0.05 & 10.68 & 0.08 & 2.74 & 0.56 & 1.1 &\\
NGC\,5866& S0  & 10.88 & 0.04 & -1.01 & 0.36 & 0.87 & 0.10 & 10.42 & 0.03 & 9.2 & 0.14 & 2.56 & 1.26 & 7.8 &\\
NGC\,7331& SAb & 11.31 & 0.11 & 0.78 & 0.15 & 0.08 & 0.32 & 11.13 & 0.04 & 10.73 & 0.09 & 2.62 & 0.64 & 1.2 &\\
\noalign{\smallskip}
\hline

\end{tabular}


\end{table*}

\begin{figure}[h]
\begin{center}
  \includegraphics[scale=0.48]{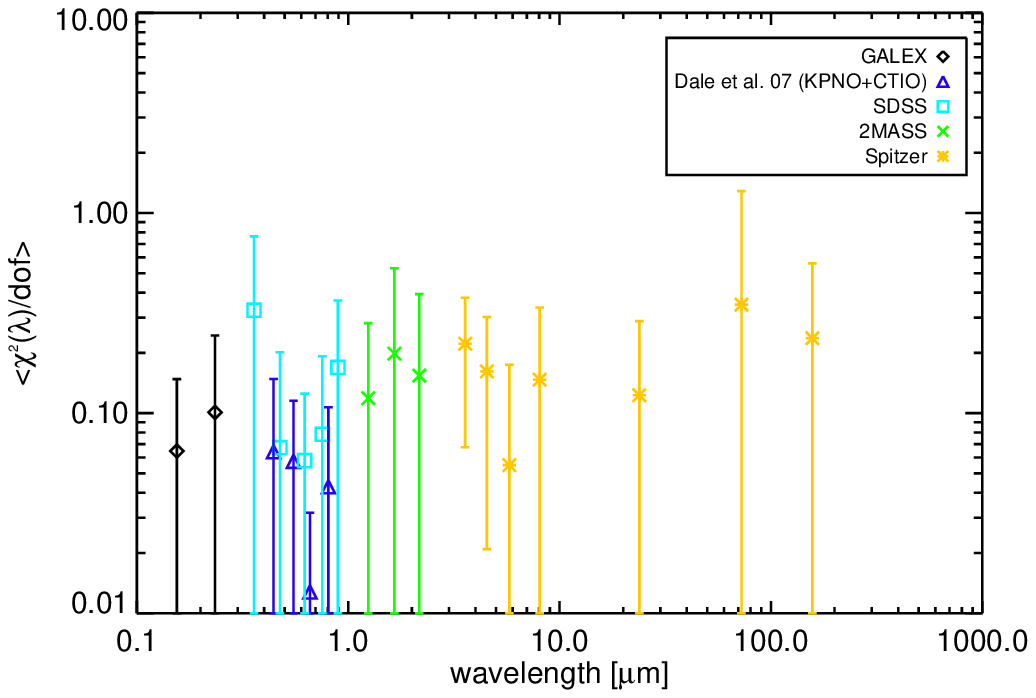} 
    \caption{The averaged frequency dependent $\mathrm{\chi^2}$ for our SINGS sample shows a general good fit for all the bands used; the large dispersion for the $70\mu$m MIPS filter is mainly due to NGC1705 and NGC5866. Error bars 
are calculated as standard deviations.}
    \label{figure7}
\end{center}
\end{figure}

\begin{table}[h!]
\begin{center}
\begin{tabular}{|c|c|c|c|}
  \hline
 \textbf{Parameters}  & \textbf{mean} & \textbf{$\sqrt\mathrm{Var}$} & \textbf{Skewness}\\
  \hline
log$_{10}~\mathrm{M_{\mathrm{star}}}$ &  -1.0  & 0.52 & -0.1  \\
log$_{10}~\mathrm{SFR}$ & 0.29 & 0.89 & 1.79 \\
log$\mathrm{t_{D4000}}$ & 1.2 & 2.8 & 2.6 \\
log$_{10}~\mathrm{L_{\mathrm{bol}}}$ & -0.45 & 0.36 & 1.17 \\
log$_{10}~\mathrm{L_{\mathrm{dust}}}$   &   -0.58 & 0.25 & 0.01   \\
$\mathrm{A_{FUV}}$ & -0.69 & 0.62 & -0.25 \\
  \hline  
  \end{tabular}
  \caption{Values of the first three moments of the $Q\equiv(\texttt{CIGALEMC}-CIGALE)/\sigma_{\texttt{CIGALEMC}}$ distributions for some parameters of interest.}
  \label{table}
\end{center}
\end{table}

\begin{figure*}[]
  \begin{center}
    \begin{tabular}{ccc}
      \resizebox{60mm}{!}{\includegraphics{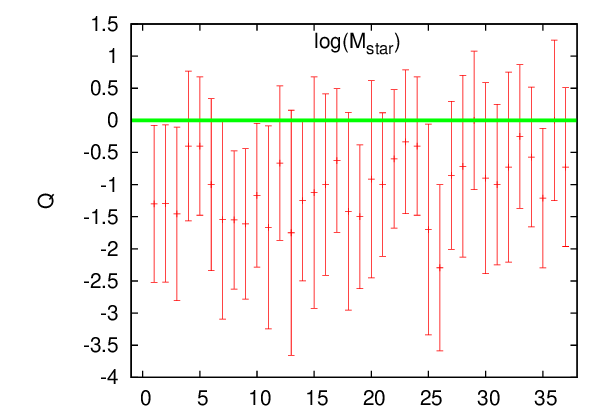}} &
      \resizebox{60mm}{!}{\includegraphics{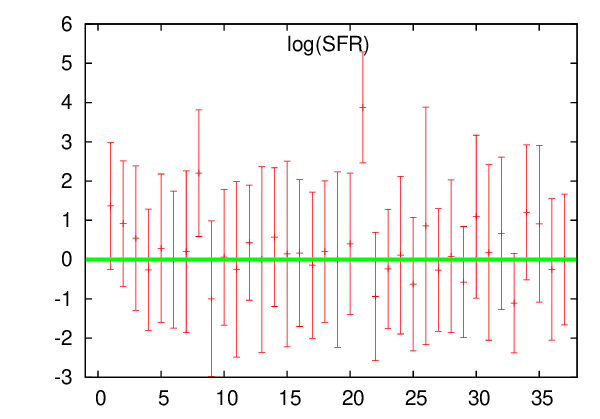}} &
      \resizebox{60mm}{!}{\includegraphics{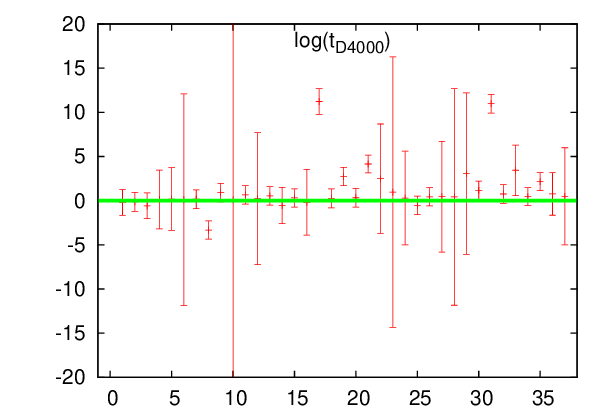}} \\
      \resizebox{60mm}{!}{\includegraphics{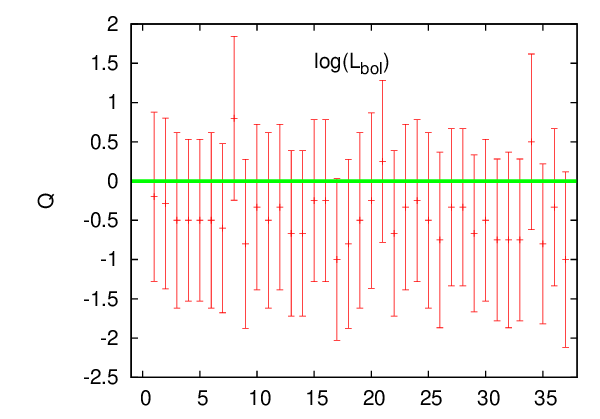}} & 
      \resizebox{60mm}{!}{\includegraphics{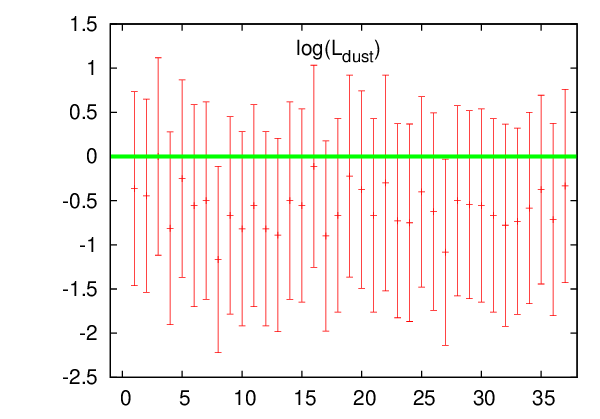}} &
      \resizebox{60mm}{!}{\includegraphics{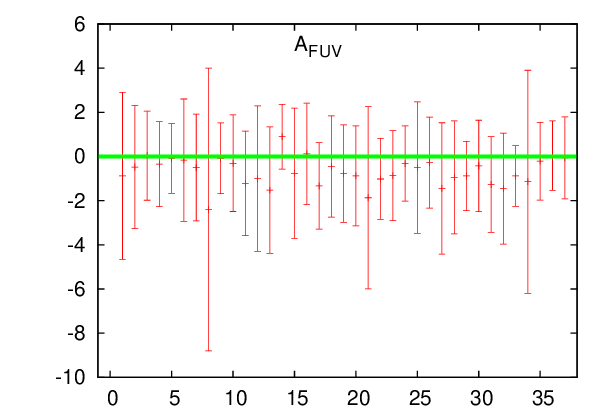}} \\
    \end{tabular}
    \caption{Ratio $\mathrm{Q\equiv(CIGALEMC-CIGALE)/\sigma_{CIGALEMC}}$ between the mean values estimated in this work and in Noll et al. (2009) for some parameters quoted in Table 3.}
    \label{plotmock}
  \end{center}
\end{figure*}

\section{Conclusions}
In this paper we have introduced a MCMC sampling method for the astrophysical 
parameter estimation from SED fitting with CIGALE. We have shown the following advantages 
of our modified \texttt{CIGALEMC} code over the usual grid-based CIGALE:
\begin{itemize}
\item its efficiency, in terms of CPU time, through the Metropolis-Hastings algorithm. 
Most of the sampling points are drawn in the region where the posterior probability is high, 
while in the grid-based approach all regions are sampled in the same way. Moreover, 
marginalized one-dimensional probability distributions for the parameters of interest are calculated by simply counting the number of samples within a binned range of parameter values, the density of sampling points being proportional 
to the posterior probability. It is hard to do the same with the usual grid approach, since the integration  
calculation scales exponentially with the number of dimensions.
The analysis of a mock sample shows that \texttt{CIGALEMC} needs 20 times less points than CIGALE to reach convergence for a given galaxy but, in general, results depend on both the number of parameters and the prior used. 
\item its accuracy; degeneracies between parameters are easily found, convergence criteria (already implemented in the code) 
ensure that statistical quantities for the parameters of interest are robustly determined and 
cross-checks through statistical analysis of mock catalogs are not necessary. 
\item its "user friendly'' characteristics. The user does not need to decide {\it a priori} the number density of samples for each region, 
trying to find a compromise between the accuracy of the results and the speed of the code: only the prior range must be chosen in advance for 
\texttt{CIGALEMC}; in fact the Metropolis-Hastings algorithm automatically samples adequately the posterior probability according to its values.
\end{itemize}
Our code will be available very soon at this web address: {\texttt http://www.oamp.fr/cigale/}.


\section{Acknowledgements}
P.S. would like to thank Denny Dale for very useful discussions. 
AA was supported by an appointment to the NASA Postdoctoral Program at
the Ames Research Center, administered by Oak Ridge Associated Universities 
through a contract with NASA. We thank Antony Lewis for providing a publicly 
available version of \texttt{cosmomc}.


\end{document}